\documentclass[12pt,preprint]{revtex4}
\usepackage{amsmath,amssymb,graphicx}
\begin{document}

%%%%%%%%%%%%%%%%%%%%%%%%%%%%%%%%%%%%%%%%%%%%%%%%%%%%%%%%%%%%%%%%%%%%%%

\title{A New Hybrid Scheme for Simulations of Highly Collisional RF-Driven Plasmas}

\author{Denis Eremin, Torben Hemke, and Thomas Mussenbrock}
\affiliation{Ruhr University Bochum, Department of Electrical Engineering 
and Information Sciences, Institute of Theoretical Electrical 
Engineering, D-44781 Bochum, Germany}

\begin{abstract}
This work describes a new 1D hybrid approach for modeling atmospheric pressure discharges featuring complex chemistry. In this approach electrons are described 
fully kinetically using Particle-In-Cell/Monte-Carlo (PIC/MCC) scheme, whereas the heavy species are modeled within a fluid description.
Validity of the popular drift-diffusion approximation is verified against a ''full'' fluid model accounting for the ion inertia
and a fully kinetic PIC/MCC code for ions as well as electrons. The fluid models require
knowledge of the momentum exchange frequency and dependence of the ion mobilities on the electric field when the ions are in equilibrium with the latter. 
To this end an auxiliary Monte-Carlo scheme is constructed. It is demonstrated that the drift-diffusion approximation can overestimate ion transport in simulations of RF-driven discharges with heavy ion species operated in the $\gamma$ mode at the atmospheric pressure or in all discharge simulations for lower pressures.
This can lead to exaggerated plasma densities and incorrect profiles provided by the drift-diffusion models. 
Therefore, the hybrid code version featuring the full ion fluid model should be favored against the more
popular drfit-diffusion model, noting that the suggested numerical scheme for the former model implies only a small additional computational cost.

\end{abstract}
\maketitle

\newpage

%%%%%%%%%%%%%%%%%%%%%%%%%%%%%%%%%%%%%%%%%%%%%%%%%%%%%%%%%%%%%%%%%%%%%%

\section{Introduction}

%%%%%%%%%%%%%%%%%%%%%%%%%%%%%%%%%%%%%%%%%%%%%%%%%%%%%%%%%%%%%%%%%%%%%%

The discharges operated under atmospheric pressure are easier to operate compared to the low-pressure discharges 
because the expensive vacuum equipment is no longer needed. However, due to their smaller size compared to
the low-pressure counterparts, the experimental diagnostics proves to be more difficult. Hence, numerical modeling
plays an important role in facilitating the understanding of processes taking place in atmospheric pressure
plasma discharges. 

%Traditionally, the plasma discharges under atmospheric pressure had been described using fluid models (e.g., \cite{sakiyama_2006}). Then
%it was realized (e.g., \cite{lay_2003}, \cite{waskoenig_2010}) that the fluid description may be inadequate even for such high-pressure discharges as
%the electrons are not in thermal equilibrium with the heavy ion or neutral species and the electron energy distribution function
%exhibits complex structure modulated in time. 

If the scale of spatial inhomogeities in a plasma discharge is much greater than the electron energy relaxation length $\lambda_{\cal E} \approx \lambda (2m_e/M)^{1/2} \gg \lambda$ (with $\lambda$ the momentum exchange mean free path
and the characteristic time scale of phenomena in interest is larger than the corresponding time scale of the energy relaxation $\tau_{\cal E} = 1/(\delta \nu)$), the spatial derivatives in the Boltzmann's equation governing the electron energy distribution function (EEDF)
can be neglected and the distribution function will only implicitly depend on the spatial inhomogeneities through the 
spatial dependence of the electric field and the plasma parameters. In this case one can use local description for
plasma behavior. However, since in most atmospheric plasmas Coulomb collisions ''maxwellizing'' the EEDF are
negligible compared to the electron-neutral collisions (usually leading to pronouncedly non-maxwellian EEDFs)
one still has to determine the EEDF and calculate the corresponding 
transport coefficients to be used in a fluid model for electrons. Typically, to this end a 0d Boltzmann solver
\cite{lay_2003, hagelaar_2005,waskoenig_2010} is used.

Whereas this popular approach is suitable for studying most atmospheric pressure discharges, 
there are situations where it turns out to be inadequate even in such highly collisional plasmas 
(e.g., \cite{iza_2005, iza_2007, kudryavtsev_2012, eremin_2015}). 
In \cite{iza_2005} it was shown that one of the indispensable ingredients in the mechanism of 
striation generation in an atmospheric dielectric barrier discharge is the non-local energy transport of the electrons.
In \cite{iza_2007} it was demonstrated for a glow discharge in helium in the $\gamma$ mode
that in contrast to the expected two-temperature EEDF it exhibits three different energy groups of electrons. In addition to 
the non-Maxwellian energetic tail on the distribution function of the electrons and the mid-energy group of electrons
common to the glow discharges operated in the $\gamma$ mode there was observed a low energy group of
electrons with an unusually low temperature contributing most to the electron density. Such a three-temperature EEDF is impossible to model with a fluid description which usually assumes a maxwellian distribution function for
electrons. The EEDF becomes even more complicated and farther from the maxwellian shape when additional
species, especially molecular gases, are added to the discharge. Furthermore, recently it has been demonstrated in \cite{eremin_2015} 
that due to the large spatial nonuniformities of the electric field atmospheric pressure plasma discharges can exhibit
distinctly nonlocal energy transport even in the $\Omega$ (ohmic) mode. These effects obviously cannot be captured by the local description.
Yet another example of nonlocal electron energy transport in the highly collisional plasmas is studied in \cite{kudryavtsev_2012},
where an efficient gas impurity detector based on a ''short'' DC discharge, which predominantly consists of the negative
glow region, is proposed. Since the transverse size of the discharge there is chosen to be smaller than the energy relaxation length,
the energetic electrons born in Penning
ionization reaction can diffuse to the discharge wall where the detector is placed, retaining a large portion of its energy
and thus producing pronounced peaks in the EEDF. The energy of such electrons is ${\cal E} = {\cal E}_* - {\cal E}_I$
with ${\cal E}_*$ the excitation energy of the working gas (chosen to be helium for the high potential energy of its metastable levels)
metastable atoms or molecules and ${\cal E}_I$ the ionization 
energy for different background gas components. Since ${\cal E}_I$ is distinct for different gas impurities, 
it is possible to identify the gas chemical composition by tracking peaks on the EEDF. If the 
electron energy transport had been local, the EEDF of the electrons leaving the discharge
would have had the maxwellian shape with no peaks on it.

Of particular importance is an appropriate description of energetic electrons. Although the number of such electrons
is too small to affect the plasma bulk dynamics, these electrons are essential for the plasma sustainment as they significantly
contribute to the particle balance through the ionization processes, which have threshold at relatively high energies. 
On the other hand, the lower energy electrons determine the bulk plasma dynamics, which in turn governs
generation of the electric field accelerating the high energy electrons.  Therefore, it is important to correctly 
reproduce the correct energy distribution function of the electrons describing all its groups. 

The details of ion energy  distribution function in the atmospheric pressure plasma discharges are, in contrast, rarely of significance as they have much smaller collisional mean free path than electrons 
and thus have the Maxwellian form in all cases of practical importance. This has motivated us to use a hybrid numerical scheme where ions are
described using the fluid approximation and the electrons are followed kinetically. Furthermore, 
most of the experimental data concerning the ion-neutral collisions are provided in the form of the reaction rates (it is much easier to find the 
differential cross-sections for the electron-atom or electron-molecule collisions) and not the energy-resolved collision cross-sections, which
further renders the kinetic description of the ions an overkill. The fluid description employed for ions also simplifies bookkeeping 
of the potentially complex chemistry taking place between the neutral and charged species and the kinetic description of the electrons ensures
an accurate energy and particle balance in the numerical description of the discharge. An additional advantage of using a fluid description for the heavy particle
species is that it requires much less computer memory compared to the kinetic description. Although this argument does not play
a significant role in 1D simulations, it can become important as long as one wants to study multidimensional models with the kinetic
part of the algorithm parallelized on graphics processing units (GPUs), as the latter usually have dedicated limited memory which
cannot be expanded.

In the present work we propose a hybrid approach, where electrons are followed 
utilizing the fully kinetic particle-in-cell technique and ions are described using a fluid model. 
Before one can trust the numerical results of a particular model, such a model has to be validated and verified by comparing its results
with either an analytical treatment or results of a well established model. Should such a comparision demonstrate any discrepancies,
they should be amenable to an explanation based on the model differences.  That is the main goal of the present work. 

%%%%%%%%%%%%%%%%%%%%%%%%%%%%%%%%%%%%%%%%%%%%%%%%%%%%%%%%%%%%%%%%%%%%%%

\section{Details of the Hybrid Code}

%%%%%%%%%%%%%%%%%%%%%%%%%%%%%%%%%%%%%%%%%%%%%%%%%%%%%%%%%%%%%%%%%%%%%%

In the present work we study plasma dynamics across the discharge channel in a RF-driven plasma (CCP) planar discharge at atmospheric pressure,
such a micro-jet discharge. 
Correspondigly, we limit ourselves to 1D spatial dimension with $z$ coordinate being
the distance from the powered electrode. The driven (grounded) electrode is located at $z=0$ ($z=L$).
Following the logic outlined in the introduction section, we adopt a kinetic description for the electron and a fluid description for 
the ion and neutral species. The resulting approach enables an efficient implementation of numerical model 
for discharges featuring a complicated chemistry (e.g., \cite{eremin_2015}). 

\subsection{Kinetic Description of Electrons}

The electron dynamics is traced numerically using 
the PIC technique \cite{birdsall_2005, hockney_1988, grigoryev_2002}, where the electron particle distribution function is
discretized on a moving Lagrangian grid with a number of markers following the characteristics of the Boltzmann's
equation, and the electric field is discretized on a stationary Eulerian grid. 
The markers used for the representation of the particle distribution function can be also regarded as
''superparticles'' representing a large number of physical particles, which are close to each other
in phase space, so that 
\begin{equation}
f_e(z,{\vec v}) = \sum\limits_j w_j W(z_j(t)-z) \delta (\vec{v}_j(t)-\vec{v}), \label{eq1.1}
\end{equation}
where $W$ is the superparticle shape function and $w_j$ is weight of the $j$th particle. In case of a field grid uniform in $z$ and
homogeneous initial plasma density (which we take to be the case in the present work), 
it is convenient to take the same weight for all superparticles, 
$ w_j = n_{e0}/N_{s0}$ with $n_{e0}$ the initial electron density and $N_{s0}$ the initial number of superparticles 
per field grid cell. This results in the following number of real particles per superparticle, 
\begin{equation}
N_{p/sp}=\frac{n_{e0}\Delta V}{N_{s0}}, \label{eq1.2}
\end{equation}
with $\Delta V$ the grid cell volume, $\Delta V = \Delta z S$, $\Delta z$ being the field grid cell size and $S$ the electrode area, respectively.

%\begin{figure}[t]
%\centering
%\includegraphics[width=15cm]{Fig1.pdf}
%\caption{}
%\end{figure}

The PIC/MCC numerical scheme describing the electrons in our implementation consists of four parts:
particle pusher, particle removal, charge density assignment,  and the Monte-Carlo collisions module. 
In the particle pusher particles are moved in phase space by integrating the Newton's equations
of particle motion under influence of the electric field. This is done using 
the explicit leapfrog scheme with the electric field interpolated from the field grid nodes to
the particle position with the interpolation function $S^\prime$ being the same as the
particle shape function $S$ in Eq.(\ref{eq1.1}), which we chose to be of the Cloud-In-Cell (CIC)
variety (e.g., \cite{birdsall_2005}). The superparticle weight remains constant during the particle pusher part of the PIC/MCC algorithm 
due to Liouville's theorem.  The pusher algorithm also checks if a particle gets reflected or generates new
particles at the borders of the computational domain due to the corresponding surface processes, such
as generation of the secondary electrons, by using corresponding probability. 
After the particle pusher is completed, it is checked if a particle
leaves the computational domain. Should it be the case, the particle is removed.
Further, one
calculates charge density of the electron component by extrapolating the superparticle charge to the
surrounding field grid nodes. 
%making sure that the extrapolation is reciprocal to the interpolation
%used in calculating the electric field at the superparticle position, which is needed to
%guarantee momentum conservation. 
Finally, a Monte-Carlo technique is used to implement the collisions
between the electron and the heavy species (charged and neutral). To this end, we exploit a variety of the "null-collision" method
described in \cite{mertmann_2011}. Rather than taking $N_e P_{max}$ number of electrons for executing a collision algorithm
with $N_e$ the total number of electrons and $P_{max} = 1 - \exp(-\nu_{max}\Delta t)$ the maximum
''null-collision'' frequency, one checks if $R<P_{max}$ with $R$ the pseudorandom number uniformly distributed in $[0,1]$  for every
particle. In electron collisions with the background neutral gas atoms or molecules the latter are treated as 
having a uniform density and any change of its density and temperature in the course of the discharge evolution
is assumed to be negligible. In contrast, when treating collisions of electrons with the other species such as metastables and
ions, the density profiles of the latter species are taken into account. When only a reaction rate is available,
%(such as for the recombination reactions between electrons and ${\rm N}^+_4$, ${\rm N}^+_2$ ions, to be discussed later)
we deduce energy dependence from the reaction rate electron temperature dependence based on the
Maxwellian ansatz and then use the resulting energy-resolved cross-section in the simulations similar to \cite{kawamura_2014}.
A proper choice of pseudorandom number generator is also important as the number of collisions in simulating such
highly collisional plasmas is large and a good pseudorandom generator should have an appropriately long period. 
For our hybrid code we used the 128 bit Xoorshift pseudorandom number generator suggested in \cite{marsaglia_2003},
which has period of $2^{128}-1$. To accelerate the computations, the kinetic PIC/MCC part of the present hybrid code was implemented
on a graphics processing unit (GPU) analagous to the GPU PIC/MCC code used in the benchmarking study \cite{turner_2013}.

The cell size $\Delta z$ was limited by the need to resolve the Debye length in order to avoid excessive numerical heating 
(in our simulations we took $\Delta z = 0.2 \lambda_{De,max}$, where $\lambda_{De,max}$ is Debye length calculated
using the maximum expected electron density $n_{e,max}$) and the time step was limited by the need to resolve
the elastic collisions (we took $\Delta t = 0.1 \nu_{max}^{-1}$ with $\nu_{max}$ the maximum value of the elastic collisions
over the computational energy domain). 

\subsection{Fluid Description of Heavy Particle Species, Full Treatment}

%For the discharge in question we kept track of six different heavy particle species:
%He$^*$, He$^*_2$, He$^+$, He$^+_2$, N$^+_2$, and N$^+_4$, where He$^*$ and He$^*_2$ are
%the helium metastable atoms and excimer molecules, respectively. 
%The list of reactions between the heavy particle species that have been
%assumed to be significant in the studied discharge is given by the reactions (R7-R15) in Table~1.  The overall set of
%reactions is analogous to those adopted in \cite{martens_2008} or \cite{lazzaroni_2012} under assumption that
%density of the background neutral N$_2$ gas is not larger than just a few percent of the He background gas density,
%thus the reactions 9 and 11 in \cite{martens_2008} are dropped with respect to the analogous reactions
%10 and 12, respectively. Also, the reaction 16 in \cite{martens_2008} describing the loss channel for the excimers
%is substituted with the more accurate similar reaction R7  from \cite{lazzaroni_2012}. One can see that the
%reaction R7 leads to a fast conversion of He$^+$ ions into He$^+_2$ ions, which in turn are rapidly converted
%into N$^+_2$ ions. Unless the N$_2$ density is very small (smaller than 10 ppm of the helium density), 
%the N$^+_2$ ions are rapidly converted to N$^+_4$ ions and the reciprocal reaction R10 of conversion of N$^+_4$
%into N$^+_2$ is much weaker than the direct reaction, so that normally N$^+_4$ nitrogen cluster ions
%is the dominant ion species in He-N$_2$ discharges \cite{martens_2008}.

In contrast to the kinetic treatment of the electron component,  the heavy species transport is followed using
a fluid model. In such a model the heavy particle density is governed by the particle continuity
equation,
\begin{equation}
\frac{\partial n_{hs}}{\partial t} + \nabla\cdot {{\bf \Gamma}_{hs}} = G_{hs} \label{eq2.1},
\end{equation}
where $hs$ is the heavy particle species subscript, ${\bf \Gamma}_{hs} = n_{hs}{\bf u}_{hs}$ is the heavy particle flux
with ${\bf u}_{hs}$ the average ordered velocity of this species, and $G_{hs}$ the net 
density change rate due to the reactions where such particles are either born or destroyed.
In general case (see, e.g., \cite{lee_2011})  the heavy particle flux is governed by equation which can be obtained by multiplying the Boltzmann's equation
by particle velocity and integrating it over velocity space, which yields
\begin{equation}
M_{hs}\left(\frac{\partial {\bf \Gamma}_{hs}}{\partial t} + \nabla\cdot \left( {\frac{{\bf \Gamma}_{hs}{\bf \Gamma}_{hs}}{n_{hs}}} \right) \right) = Z_{hs} e n_{hs} {\bf E} - \nabla (n_{hs} T_{hs}) - \nu_{hs,N}M_{hs}{\bf \Gamma}_{hs} \label{eq2.2}
\end{equation}
with ${\bf E}$ the electric field, $Z_{hs}$ the species charge number, $M_{hs}$ its mass, $T_{hs}$ the species temperature, $\nu_{hs,N}$ the momentum exchange frequency of the species with the neutral background gas $N$, and the expression in the brackets of the second term on the left hand side is a dyadic tensor. 
In the atmospheric pressure plasma discharges the left hand side of Eq.~(\ref{eq2.2}) is normally neglected and the ion fluxes are calculated from the right hand
side of this equation (drift-diffusion approximation, see the next section). However, it is pointed out in \cite{lee_2011} that the right hand side of Eq.~(\ref{eq2.2})
vanishes only after the equilibrium drift velocity is achieved and it takes several nanoseconds for a heavy ion to accomplish it. A large ion flux gradient, 
which can arise in a highly collisional discharge, for example, due to the intense ionizing electron avalanche in the $\gamma$ mode, can cause the second
term on the l.h.s. of Eq.~(\ref{eq2.2}) to be of significance as well. Since the drift-diffusion approximation is also frequently used for lower pressure discharges
(up to 100 mTorr), the explicit time derivative in Eq.~(\ref{eq2.2}) can clearly become comparable with the last term in this parameter range. All this suggests that the left hand side of the latter equation can be substantial and it is interesting to verify how close the results of the full model utilizing Eq.~(\ref{eq2.2}) are to the results of the popular drift-diffusion approximation under different conditions.

Finally, we note that the energy transport equation is usually omitted for the heavy species since they come into the thermodynamical equilibrium with the background
neutral gas really fast due to the efficient collisional energy exchange in contrast to the electrons. Under this assumption the temperature
is usually approximated as $T_N$ for the neutral species and by Eq.~(\ref{eq2.8}) for the ion species.

A convenient numerical scheme for solving Eqs.~(\ref{eq2.2}) and (\ref{eq2.1}) employs the leapfrog time intergration approach,
where discretization of the density $n$ is performed at integral time levels, and the flux ${\bf \Gamma}$ is taken at time levels half a time step apart. 
In the particular scheme that we have used in this study one first calculates the particle flux at the next time level, so that in the one dimensional
case the corresponding equations read
\begin{equation}
\begin{array}{lll}
\Gamma_j^{n+1/2} \left(1 + \frac{\nu_j^n \Delta t}{2} \right) &+& \frac{\Delta t}{2\Delta z} \left( \frac{\Gamma^{n+1/2}_{j+1}\Gamma^{n-1/2}_{j+1}}{n_{j+1}^n}-\frac{\Gamma^{n+1/2}_{j-1}\Gamma^{n-1/2}_{j-1}}{n^n_{j-1}} \right)  \\
&=& \Gamma^{n-1/2}_j \left(1- \frac{\nu^n_j \Delta t}{2} \right) + \frac{Z e n^n_j  E^n \Delta t}{M} - \frac{\Delta t}{2\Delta z M} (n_{j+1}^n T_{j+1}^n-n_{j-1}^n T_{j-1}^n)
\end{array}
\label{eq2.3}
\end{equation}
for $j=1..N-2$, which can be easily found by integrating Eq.~(\ref{eq2.2}) from $z_{j-1/2} = (j-1/2)\Delta z$ to $z_{j+1/2}=(j+1/2)\Delta z$. The 
equations at the boundary nodes are obtained by the same integration procedure using the corresponding intervals $[z_0, z_{1/2}]$ and $[z_{N-3/2},z_{N-1}]$, 
respectively. The procedure being straightforward, we omit the results. Note that the advection term on the l.h.s. of these equations is discretized semi-implicitly, which helps to keep the equations linear and tridiagonal. Then, they can be easily solved, for example, by using the Thomas' method. Generally speaking, the momentum exchange frequency depends on velocity
$u=\Gamma/n$, which is spatially nonuniform. 
To avoid the need for an iterative solver we use $u^n = \Gamma^{n-1/2}/n^n$, which reduces the order of the numerical scheme, but
we find in the benchmark comparisons with PIC simulations the resulting accuracy to be still sufficient due to the small time-step caused by the
need to resolve the collisions in the PIC/MCC part of the algorithm describing electrons. A higher oder alternative would be to solve Eqs.~(\ref{eq2.3}) and (\ref{eq2.4})
iteratively, which is more computationally expensive. Although the diffusion term proved to
be small in all the cases we considered, we still retained it, albeit using a simplified expression from the drift-diffusion approximation given in Eqs.~(\ref{eq2.7}) and (\ref{eq2.8}). To determine the velocity-dependent momentum exchange frequency $\nu$ in Eq.~(\ref{eq2.3})
we have constructed an auxiliary Monte-Carlo code, which calculates it  based on the energy resolved collisional cross-sections. A detailed desciption of the latter code is given in the next subsection. 

After the particle flux at the next half-integer time level is found, one can obtain the value of the density at the next integer time level from the particle continuity
equation,
 \begin{equation}
n^{n+1} = n^{n} - \frac{\Delta t}{2\Delta z}\left(\Gamma^{n+1/2}_{j+1}-\Gamma^{n+1/2}_{j-1}\right) + \Delta t G^{n+1/2} \label{eq2.4},
\end{equation}
for $j=1..N-2$, which is derived by integrating the particle continuity equation over $[z_{j-1/2},z_{j+1/2}]$. Equations for the bondary points
$j=0$ and $j=N-1$ are obtained similarly by integrating over the corresponding intervals  $[z_0, z_{1/2}]$ and $[z_{N-3/2},z_{N-1}]$
and are not shown here. For the sake of simplicity in our numerical implementation we substitute $G^{n+1/2}$ with $G^n$, which 
degrades accuracy of the scheme. However, by comparison with the PIC simulations we conclude that the resulting accuracy is still sufficient
not to cause a significant deviation from the fully kinetic calculations.

\subsection{Fluid Description of Heavy Particle Species, Drift-Diffusion Approximation}

In the frequently used drift-diffusion approximation the particle fluxes are assumed to have reached a quasi-stationary value, 
so that the left hand side of Eq.(\ref{eq2.2}) is neglected. In this case the particle flux
is approximated as 
\begin{equation}
{\bf \Gamma}_{hs} = Z_{hs}e n_{hs}\mu_{hs} {\bf E} - D_{hs}\nabla n_{hs} \label{eq2.5}
\end{equation}
with 
\begin{equation}
\mu_{hs} = \frac{Z_{hs}e}{M_{hs}\nu_{hs,N}} \label{eq2.6}
\end{equation}
the heavy particle mobility (note that it equals zero for neutral species), and $D_{hs}$ the heavy particle diffusion rate.

The diffusion rate accounting for the impact of electric field
on the ion diffusivity is obtained by using the generalized Einstein's relation \cite{ellis_1976}, 
\begin{equation}
D_i = \frac{\mu_i T_i}{q_i} \label{eq2.7}
\end{equation}
with
\begin{equation}
T_i = T_{\rm N} + \frac{m_i + m_{\rm N}}{5m_i + 3m_{\rm N}} m_{\rm N}(\mu_i |E|)^2. \label{eq2.8}
\end{equation}
%and assuming that collisions with the helium working gas dominate due to its large density.

Using the expression for particle flux in the drift-diffusion approximation given in Eq.~(\ref{eq2.5}) one can 
obtain particle density from Eq.~(\ref{eq2.4}). This equation can be solved either explicitly
(by taking the particle flux values not at the time level $n+1/2$, but $n$), in which case the boundary condition is not needed, 
or, for example, semi-implicitly (if the original form of Eq.~(\ref{eq2.4}) is retained. 
In the latter case the boundary conditions come from the kinetic limitation on the particle flux to the wall for the 
neutral species, $({\bf\Gamma}_n\cdot {\bf n})(z=0,L) = \frac{n}{4}\sqrt{8 T_n/\pi m_n}$ with ${\bf n}$ the normal vector
to the corresponding electrode, and from the assumption of the electric field dominated particle flux to the wall for the ions, 
$({\bf\Gamma}_i\cdot {\bf n})(z=0,L) = \alpha\mu_i n_i {\bf E}$, with $\alpha$ the switching function, taking value
$1$ when ${\bf E}\cdot{\bf n}>0$ and $0$ when ${\bf E}\cdot{\bf n}\leq 0$, respectively (e.g. \cite{sakiyama_2006}). 

It is essential to correctly determine the momentum exchange frequency $\nu_{hs,N}$ in Eq.~(\ref{eq2.3}) 
and the mobilities in Eq.~(\ref{eq2.5}) (note that one of these quantities can be calculated from the other utilizing  Eq.~(\ref{eq2.6}))
caused by collisions of ion species with the helium background gas.
Although ion mobilities are often assumed to be constant (see, e.g., \cite{sakiyama_2006}), a better approximation
attempted from the physics considerations yielding dependence of the mobilities on the electric field demonstrates that
such a dependence can matter \cite{greb_2013}. Following the latter argument,
in the present work we propose to calculate mobilities
and the corresponding momentum exchange frequency with help of the following auxiliary Monte-Carlo code.

The code resembles very much a PIC/MCC code where only one of the ion species is traced. 
The particles are evolved in time under action of a prescribed constant electric field $E_0$ 
and collisions with the background neutral gas having a temperature $T_N$. The collisions are
modeled using the Monte-Carlo method with the same energy resolved cross-sections as in the full PIC code.
At the initial moment the particle velocities are sampled from a maxwellian distribution with an initial temperature $T_{i0}$.
The computational domain is assumed to have periodic boundaries. After several nanoseconds the ions acquire 
the stationary drift velocity
(see Figure 6 in \cite{lee_2011} and the accompanying discussion in that reference),
which can be determined by calculating the total mean velocity of the particle ensemble. Carrying out
this procedure for a number of the electric field values, one can construct a look-up table or an
analytic fit of mobility (calculated as $u/E_0$, where $u$ is the drift velocity) and momentum exchange frequency
(calculated from Eq.~(\ref{eq2.6})) versus the electric field and the drift velocity, respectively.
Following the suggestion in \cite{phelps_1994}, we use both in the full PIC and the Monte-Carlo code
for the ion  isotropic scattering the cross-section equal to $Q_m = Q_i + 2Q_b$, where
$Q_i$ is the elastic isotropic cross-section and $Q_b$ is the elastic backward scattering 
cross-section. 

Note that this technique is more general than the one suggested in \cite{greb_2013}. It enables an accurate calculation of
the momentum exchange frequencies and mobilities for any species, for which energy-resolved collision cross-sections are
known, and under any conditions of practical interest (for example, when the colliding species have different 
velocity distribution functions with comparable but distinct characteristic energy/temperature). Calculation
of the corresponding collision frequencies and mobilities has to be performed only once for a given pressure of
the background helium gas and thus does not cause any computational overhead in the hybrid code simulations.

\subsection{Coupling Between the Electron and Ion Models}

Coupling of the kinetic electron and the fluid ion models occurs through the net charge density in the Poisson's equation,
which includes both electron and ion densities and through any reaction involving both electrons and ions.
Among the typical reactions are the secondary electron emission caused by ions impinging
on the electrode surfaces, production of electrons and positive ions in the electron impact, Penning and the metastable pooling ionization reactions, 
and recombination reactions of electrons with positive ions.

Once the charge densities are calculated, the Poisson's equation has to be solved. 
In order to achieve a needed regime of operation for the discharge in numerical
simulations one has to limit the current flowing through the discharge. If it is not done
(for example, if a fixed voltage amplitude RF source is used in the simulations),
the number of superparticles and hence the particle density tend to diverge with time as the modeled discharge spontaneously
goes into the $\gamma$ regime with an ever increasing current. In order to limit
the total current in the simulations one can chose either to use a fixed amplitude
currrent source or, if a fixed amplitude voltage source is used, either to use an external resistance or to
limit the power absorbed in the discharge by adjusting the voltage amplitude to meet the prescribed power. The latter approach
seems to correspond to experimental observations better, the latter showing
generation of additional harmonics rather in the measured current than in the voltage signal \cite{waskoenig_2010}.
To adapt the voltage ampltide for matching the prescribed power we have implemented
the following simple algorithm. Starting from an initial guess for the voltage amplitude $U^0$,
during each RF period the voltage ampltiude remains fixed and the period-averaged power absorbed by the discharge 
during the $N$th period is calculated as 
$P^{N}_{abs} = \frac{S}{T} \int^{T(N+1)}_{TN} V (j + \epsilon_0 \partial E/\partial t)dt$,
where $S$ is the electrode area, $V$ is the voltage, $T$ is the RF period, $j$ and $\epsilon_0 \partial E/\partial t$ are the
conduction current and the displacement current densities at the driven electrode.
The new value for the voltage amplitude during the $(N+1)$th RF period is calculated as
\begin{equation}
U^{N+1} = U^N \left(1-\alpha \frac{(P_{abs}^N-P_0)}{P_0} \right) \label{eq2.9}
\end{equation}
with $P_0$ the prescribed power and $\alpha$ some numerical parameter, which can be chosen to be small for a smoother
or large for a faster adjustment, respectively. It can be also be chosen to be time-adaptive, for example,
being larger at the beginning of the simulation and smaller as the simulation proceeds.

The secondary electron emission algorithm is implemented by employing the following technique. 
The number of heavy particles (ions or metastables) $\Delta N_i$ hitting an electrode during a time interval $\Delta t$ equals to ${\Gamma}_i S\Delta t$. 
Each such heavy particle can produce a secondary electron with probabilty equal to 
$\gamma_i$, which results in an average number of secondary electrons produced during a time step
equal to $N_{sec.el} = \gamma_i \Delta N_i$. This translates to $N_{sec.el}/N_{p/sp}$ superparticles with $N_{p/sp}$ 
defined in Eq.~(\ref{eq1.2}). Thus, during each time step one produces  $int \left(N_{sec.el}/N_{p/sp}\right)$
(with $int$ representing the integral part of a number) electrons and performs a comparison of the $R$ uniformly 
distributed pseudorandom number with $N_{sec.el}/N_{p/sp} - int \left(N_{sec.el}/N_{p/sp}\right)$. If the former
is smaller than the latter, one additional secondary electron superparticle is produced. Such a Monte-Carlo technique
ensures correct number of secondary electrons generated on average. 

The electron impact ionization is treated according to the modified null-method described in \cite{mertmann_2011}. 
When an impact ionization or a metastable excitation event occurs, a corresponding location of the newly
produced ions or metastable species are recorded and the corresponding ion or metastable density
is incremented by $N_{p/sp}/\Delta V$ with $\Delta V$ the grid cell volume
along with creation of a new electron superparticle. Velocity of the ejected electrode is determined
by a random scattering on a sphere in velocity space with radius corresponding to the energy 
of the ejected electron \cite{vahedi_1995}.

Ionization events due to the Penning or the metastable pooling reactions are treated in a similar fashion, but
with the number of electron superparticles corrected with regard to the corresponding reaction.
For example, for the Penning ionization such number is equal to $\Delta N_e/N_{p/sp}$, where
number of electrons produced during a time step is equal to
$\Delta N_e = R n_*  n_N \Delta t \Delta z S $ with $R$ the corresponding reaction rate,
$n_*$ the density of the corresponding metastable species and $n_N$ the density of the background
neutral gas ionized by the metastable species. As in the case of the impact ionization, velocity
of the electron created in such a process is to be found by scattering on
a sphere in velocity space with radius calculated from the corresponding energy (different
for each of the processes).

The recombination reactions between electrons and positive ions are accounted for numerically
as follows. First, the maximum collision probability during a time step is calculated as $P_{max} =  1 - \exp(-\nu_{max}\Delta t)$
with $\nu_{max} = \max\limits_z{n_i(z)}\cdot \max\limits_{\cal E} (\sigma({\cal E}) v)$. Then, the rest of the null
collision procedure goes as the previously mentioned modified null-collision method of \cite{mertmann_2011} 
with a correction respecting the nonuniform ion density profile when calculating the actual collision probability. 
When a recombination event occurs, the corresponding electron superparticle is removed from the
computational domain and the corresponding ion density is decremented by $N_{p/sp}/\Delta V$.

%Fig.~1 illustrates the corresponding total numerical scheme of the hybrid code.

%%%%%%%%%%%%%%%%%%%%%%%%%%%%%%%%%%%%%%%%%%%%%%%%%%%%%%%%%%%%%%%%%%%%%%

\section{Validation of the Hybrid Code}

%%%%%%%%%%%%%%%%%%%%%%%%%%%%%%%%%%%%%%%%%%%%%%%%%%%%%%%%%%%%%%%%%%%%%%

\begin{table}
\caption{Electron-neutral collisions}
\begin{tabular}{c l l}
\hline
(R1)   & e + He $\to$ e + He & $\sigma({\cal E})$ \, \cite{verboncoeur_1993} \\
(R2)   & e + He $\to$ e + e + He$^+$ & $\sigma({\cal E})$ \, \cite{verboncoeur_1993} \\
(R3)   & e + He $\to$ e + He$^*$ & $\sigma({\cal E})$ \,  \cite{verboncoeur_1993} \\
(R4)   & He$^+$ + He $\to$ He$^+$ + He &  \,\,\,  $Q_i({\cal E}) + 2Q_b({\cal E})$ \,  \cite{jila}, see also text
\end{tabular} \label{t1}
\end{table}

The approach described in the previous section enables an efficient implementation of numerical model 
for discharges posessing a complicated chemistry. The goal of the present
work is, however, to validate the approach itself and to demonstrate its applicability to adequate modeling of the highly
collisional plasma discharges. To this end in the next section we compare the simulation results of the hybrid
code with the simulation results of the GPU PIC/MCC code. The latter has been verified in the benchmark study of \cite{turner_2013}, 
but this time it has been ran assuming the atmospheric pressure in a pure He discharge using a simple reduced chemistry set similar to the one used
in \cite{iza_2005} and listed in Table~\ref{t1}. Such a chemistry set does not represent the actual physics taking place
in such discharges because metastable atoms and excimer molecules are neglected, whereas they
usually dominate the electron production through the pooling reactions. However, the simplified chemistry
allows to focus on validating if the hybrid approach properly describes the particle, momentum, and energy transport,
which should match those of the PIC code making minimum assumptions and thus describing the
physics in such discharges comprehensively. Hence, in this reduced set only electrons and He$^+$ ions 
are tracked. 

For a pure helium discharge we took the energy resolved elastic cross-sections for the collisions
between the electrons and the He background neutral gas from the XPDP1 code \cite{verboncoeur_1993} (for the benchmarking
purposes a particular choice of these cross-sections does not play a significant role, since the electron component
is modeled using the same approach and the same parameters, including the cross-sections, both in the hybrid code and in the fully kinetic GPU PIC/MCC code), and for
the collisions of He$^+$ ions with the He neutrals from \cite{jila}, 
$Q_i({\cal E}) = 7.63\times 10^{-20}/ \sqrt{\cal E} \,\, [{\rm m}^2]$ and $Q_b({\cal E}) = 10^{-19}/ [({\cal E}/1000)^{0.15} (1+{\cal E}/1000)^{0.25} (1+5/{\cal E})^{0.15}] \,\, [{\rm m}^2]$ with ${\cal E}$ the relative energy in eV calculated in the center of mass $M_r v_{rel}^2/2 \approx M_{He} v_{rel}^2/4$, where $M_r$ is the reduced mass of the collision partners and $v_{rel}$ is their relative velocity. Utilizing these cross-sections, the Monte-Carlo code described in the previous section 
yielded mobilities for the He$^+$ ions in a He gas under atmospheric pressure, which we  approximated by the analytical fits listed in the Table~\ref{t2}.
\begin{table}
\caption{Mobilities, $10^{-3} \,\, \left[{\rm m^2/(V\, s})\right]$ }
\begin{tabular}{l l l}
\hline
$1.18  - 0.185 (|E|/E_0)$   &,&  $|E|<E_0 = 1.47 \times 10^6 \,\, {\rm V/m}$ \\
$0.94 (E_0/|E|)^{0.3}$ &,& $|E|\ge E_0$
\end{tabular} \label{t2}
\end{table}
\begin{table}
\caption{Momentum Exchange Frequency, $10^{10} \,\, \left[{\rm s^{-1}}\right]$ }
\begin{tabular}{l l l}
\hline
$2.07  + 0.88 (|u|/u_0)^{1.89}$   &,&  $|u|<u_0 = 2\times 10^3 \,\, {\rm m/s}$ \\
$1.54 + 1.36 (|u|/u_0)$ &,& $|u|\ge u_0$
\end{tabular} \label{t3}
\end{table}
 A better fit can be made to match expected the high field asymptotics, $\mu \propto E^{-0.5}$ (see the discussion in \cite{greb_2013}),
more accurately, yet we have found that the simple formula given in Table~\ref{t2} gives a good agreement for $E<5\times 10^6 \,\, {\rm V/m}$,
which is the range of the electric field values observed in the simulations. Similarly, the analytical fit for the momentum exchange frequency in the elastic
scattering between He$^+$ ions and He atoms is listed in Table~\ref{t3}. The secondary electron emission coefficients were calculated using the empirical formula given in \cite{raizer_1991} (the reference quotes 50\% accuracy for this formula), $\gamma_s \approx 0.016 (E_{iz} - 2 E_\phi)$ with $E_{iz}$ the ionization potential of the incident ion and $E_\phi \approx 4.5\,\,{\rm eV}$ the work function. This yields $\gamma \approx 0.25$ for the helium ions.

\begin{figure}[t]
\centering
\includegraphics[width=15cm]{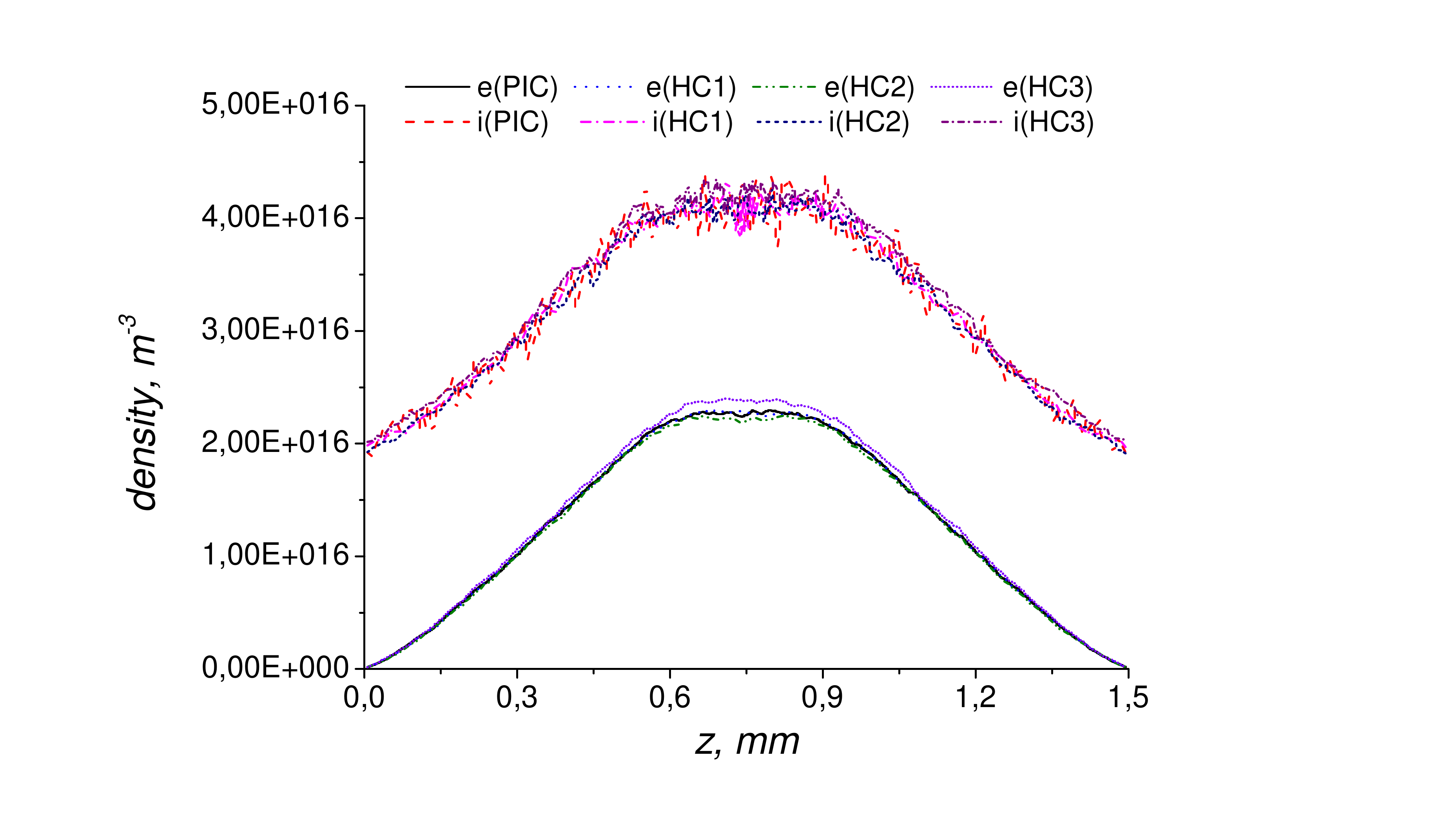}
\caption{Shown are RF period-averaged density profiles for the Case I ($\Omega$ mode) calculated with the PIC code and the different versions of the hybrid
code, see text.}
\end{figure}

For the first set of benchmarks testing the hybrid code against the particle-in-cell code for the atmospheric pressure capacitively coupled plasma discharge
we have chosen to study two different regimes of the discharge operation, an ohmically heated discharge ($\Omega$ mode \cite{hemke_2013}) with $P_0/S = 10^4 \,\, {\rm W/m^2}$ and $l=1.5\,\,{\rm mm}$ (Case I) and a discharge dominated by the ionizing avalanches produced by the 
secondary electrons accelerated by strong electric field in a discharge with the electrode distance $l=75\,\,\mu{\rm m}$ comparable to its sheath width, so that
we call it a ''short'' $\gamma$ discharge (Case II). For the latter discharge the power density is considerably higher, we studied a case with $P_0/S = 2.4 \times 10^6 \,\, {\rm W/m^2}$. 

Fig.~1 shows simulation results of the Case I for the RF period-averaged particle denisty profiles provided by the PIC/MCC code and three different versions of the hybrid code. The first version of the hybrid code (''HC1'') employs the full fluid model described in the previous section, which accounts for the ion inertia by solving Eq.~(\ref{eq2.3}) and uses the analytic fit for the momentum exchange frequency given in Talbe~\ref{t1},  determined with help of the Monte-Carlo code described in the previous section. The second version of the hybrid code (''HC2'') uses the drift-diffusion approximation with the analytic fit for the  He$^+$ 
mobilities in He given in Table~\ref{t2}, also obtained from the results of the auxiliary Monte-Carlo code. Finally, the third hybrid code version (''HC3'') uses the drift-diffusion approximation described in the previous section and a simple constant approximation for the mobility value (taken from the same analytic fit  at zero electric field, which incidentally turns out to be very close to the value adopted in \cite{sakiyama_2006}). The comparison between simulation data reveals that this regime 
is quite accurately described by all the considered fluid models. As expected, the largest deviation from the kinetic results is demonstrated by the constant-mobility
model (HC3), yet its results are still very close to prediction of the PIC/MCC code.

% One can clearly see the striking difference
%between the density profiles provided by the hybrid codes with the drift-diffusion model for the electron transport and the PIC/MCC code. This difference
%is likely to be attributed to the fact that due to substantial inertia it takes a few nanoseconds even for the light He$^+$ ions to come into equilibrium with the electric %field
%and achieve their stationary drift velocities, whereas the drift-diffusion approximation assumes instantaneous equilibrium for the ions. The ion transport impeded
%by the ion inertia also affects the electron transport coupled to the ion transport through the electric field, which seems to reduce the net conduction current
%and, as a result, plasma heating. Therefore, the kinetic PIC/MCC code exhibits significantly lower electron density in the bulk plasma compared to that calculated
%with the drift-diffusion model. In contrast, the hybrid code version taking into account the ion inertia yields electron density being very close to the fully kinetic result.

\begin{figure}[t]
\centering
\includegraphics[width=12cm]{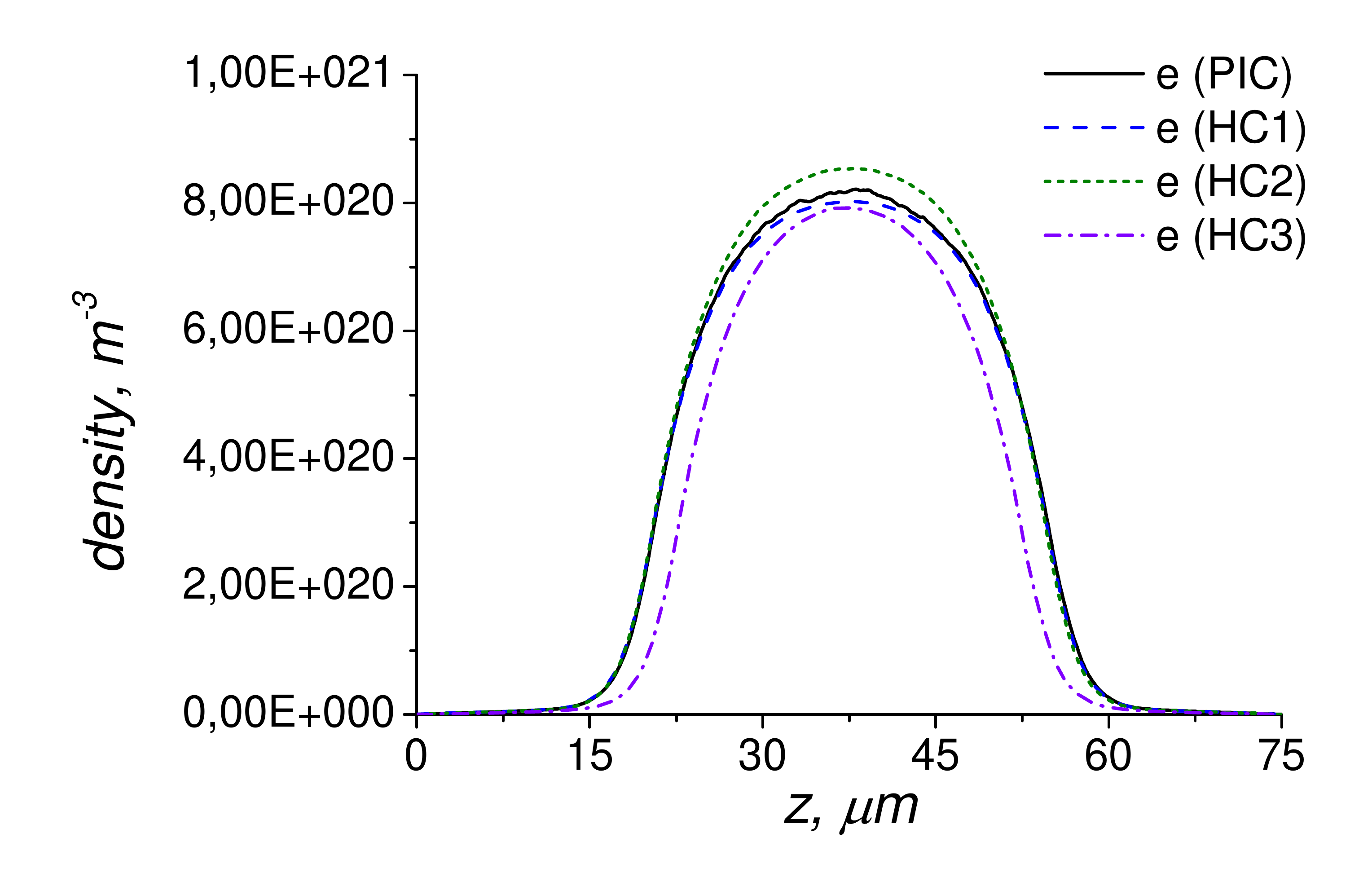}
\caption{The RF period-averaged density profiles for the Case II ($\gamma$ mode, ions with Xe mass) obtained by the PIC/MCC code and different versions of the hybrid code, see text.}
\end{figure}

Fig.~2 shows results of the benchmark of the PIC/MCC code and the same versions of the hybrid code as described above used for parameters of the Case~II ($\gamma$ mode, in which the discharge is sustained by the ion-induced secondary electron emission and the ionization avalanches caused by the secondary electrons 
accelerated by the electric field).  In this regime plasma density is much higher than in the $\Omega$ mode and the bulk plasma is quasineutral on average. Therefore, 
we have plotted only the RF period-averaged electron densities and omitted the ion densities, which are equal to the electron densities in the major part of the discharge, in order to simplify the figure. One can see that the drift-diffusion approximation gives a relatively good agreement with the prediction of the PIC code and the full fluid model used in the HC1 version of the hybrid code. The three profiles (PIC, HC1, and HC2) are very close over almost entire gap and differ only slightly in the center region. In contrast, the hybrid code version using the constant mobility (the zero electric field value) demonstrates an average deviation from the kinetic result. The difference between the kinetic result and the result of the constant mobility drift-diffusion model can be attributed to the fact that electric field in the studied example is much higher than in the $\Omega$ mode and thus dependence of the helium ion mobility on the electric field given in Table~\ref{t2}
starts to play a role. Still, the discrepancy between the peak values obtained with the PIC and the HC3 codes are less than 4\% and thus are not significant. 

\begin{table}
\caption{Mobilities, $10^{-4} \,\, \left[{\rm m^2/(V\, s})\right]$ }
\begin{tabular}{l l l}
\hline
$3.05 - 0.75(|E|/E_0)$   &,&  $|E|<E_0 = 2 \times 10^6 \,\, {\rm V/m}$ \\
$2.3(E_0/|E|)^{0.37} $ &,& $|E|\ge E_0$
\end{tabular} \label{t4}
\end{table}
\begin{table}
\caption{Momentum Exchange Frequency, $10^9 \,\, \left[{\rm s^{-1}}\right]$ }
\begin{tabular}{l l l}
\hline
$6 + 7\times 10^{-6} |u|^{2.04}$   &,&  $|u|<u_0 = 4\times 10^2 \,\, {\rm m/s}$ \\
$4.5 + 27|u|/u_0$ &,& $|u|\ge u_0$
\end{tabular} \label{t5}
\end{table}

\begin{figure}[t]
\centering
\includegraphics[width=12cm]{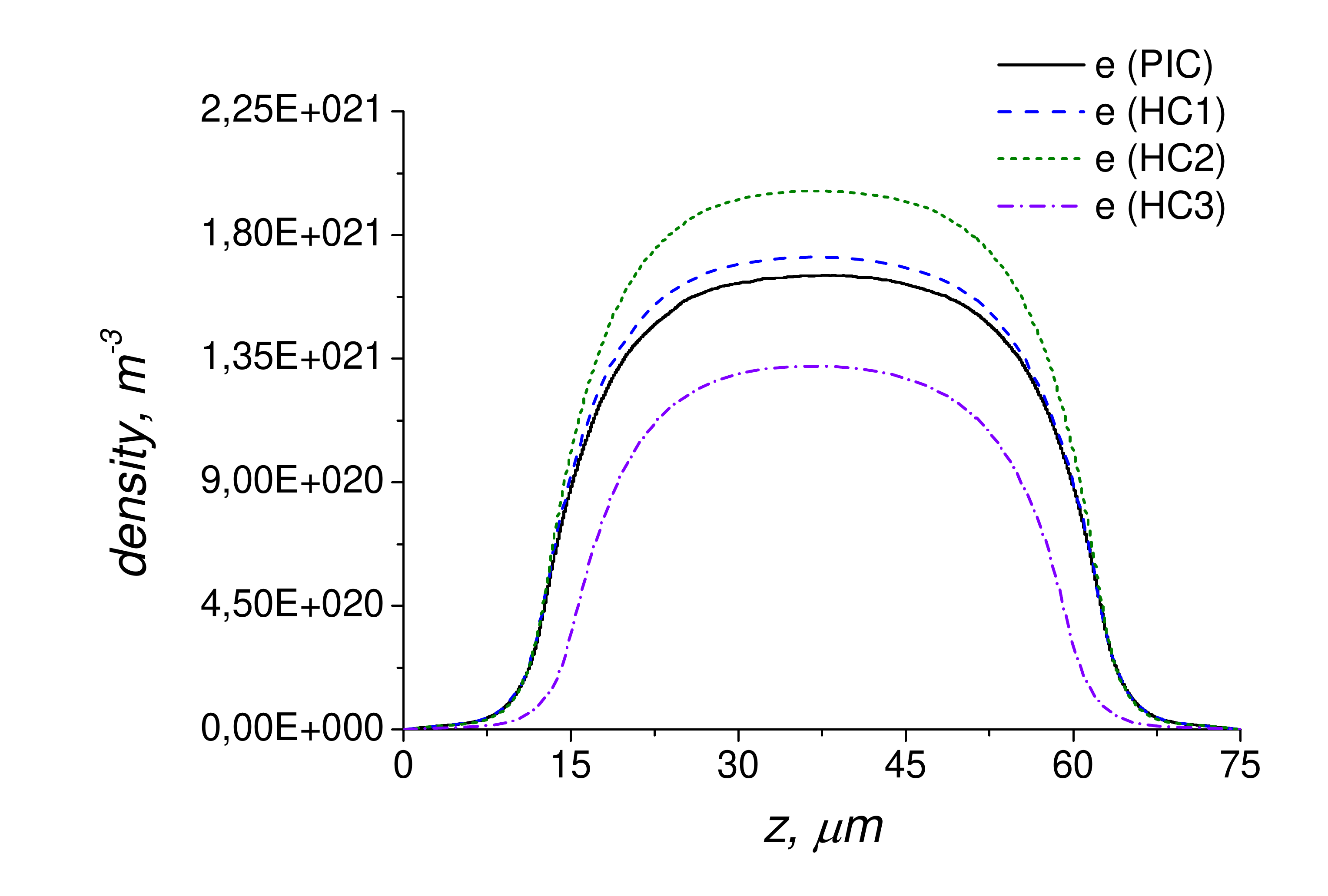}
\caption{The RF period-averaged density profiles for the Case III ($\gamma$ mode) obtained by the PIC/MCC code and different versions of the hybrid code, see text.}
\end{figure}

In \cite{lee_2011} it was suggested that heavier ions should require more time to reach the stationary drift velocity under influence of electric field and collisions with the background gas. To see how ion mass affects the accuracy of the fluid models used in the hybrid code we have performed the following test. We have used the same ion-neutral collision crosssections as for the He$^+$ - He collisions given in the Table~\ref{t1} before, but this time we have increased the ion mass to match that of xenon.
The Monte-Carlo code used to calculate the momentum exchange frequency and mobilities has provided data, which we fitted analytically as shown in Tables~\ref{t4} and \ref{t5}. The corresponding simulations conducted with parameters similar to the Case I ($\Omega$ mode) have shown very small discrepancy between the models and is not shown here. However, an analog of the Case II (''short'' $\gamma$ discharge) simulated with the heavier ions (henceforth called Case III) indeed exhibits dependence of the resulting RF-averaged electron density profiles on a particular model (see Fig.~3). Whereas the profiles calculated with the PIC/MCC code and the full fluid model (HC1) are still quite close to each other, the drift-diffusion models with variable mobility (HC2) and constant mobility (HC3) demonstrate a 20\% digression from the kinetic result. Interestingly, the latter models yield different signs of the deviation from the kinetic result.

\begin{table}
\caption{Mobilities, $\,\, \left[{\rm m^2/(V\, s})\right]$ }
\begin{tabular}{l l l}
\hline
$2.2 - 0.78(|E|/E_0)$   &,&  $|E|<E_0 = 2.8 \times 10^3 \,\, {\rm V/m}$ \\
$1.42 (E_0/E)^{0.41} $ &,& $|E|\ge E_0$
\end{tabular} \label{t6}
\end{table}
\begin{table}
\caption{Momentum Exchange Frequency, $10^6 \,\, \left[{\rm s^{-1}}\right]$ }
\begin{tabular}{lll }
\hline
$8 + 0.49|u|/2.e4$ &,& $u<10^5 \,\, {\rm m/s}$
\end{tabular} \label{t7}
\end{table}

\begin{figure}[t]
\centering
\includegraphics[width=12cm]{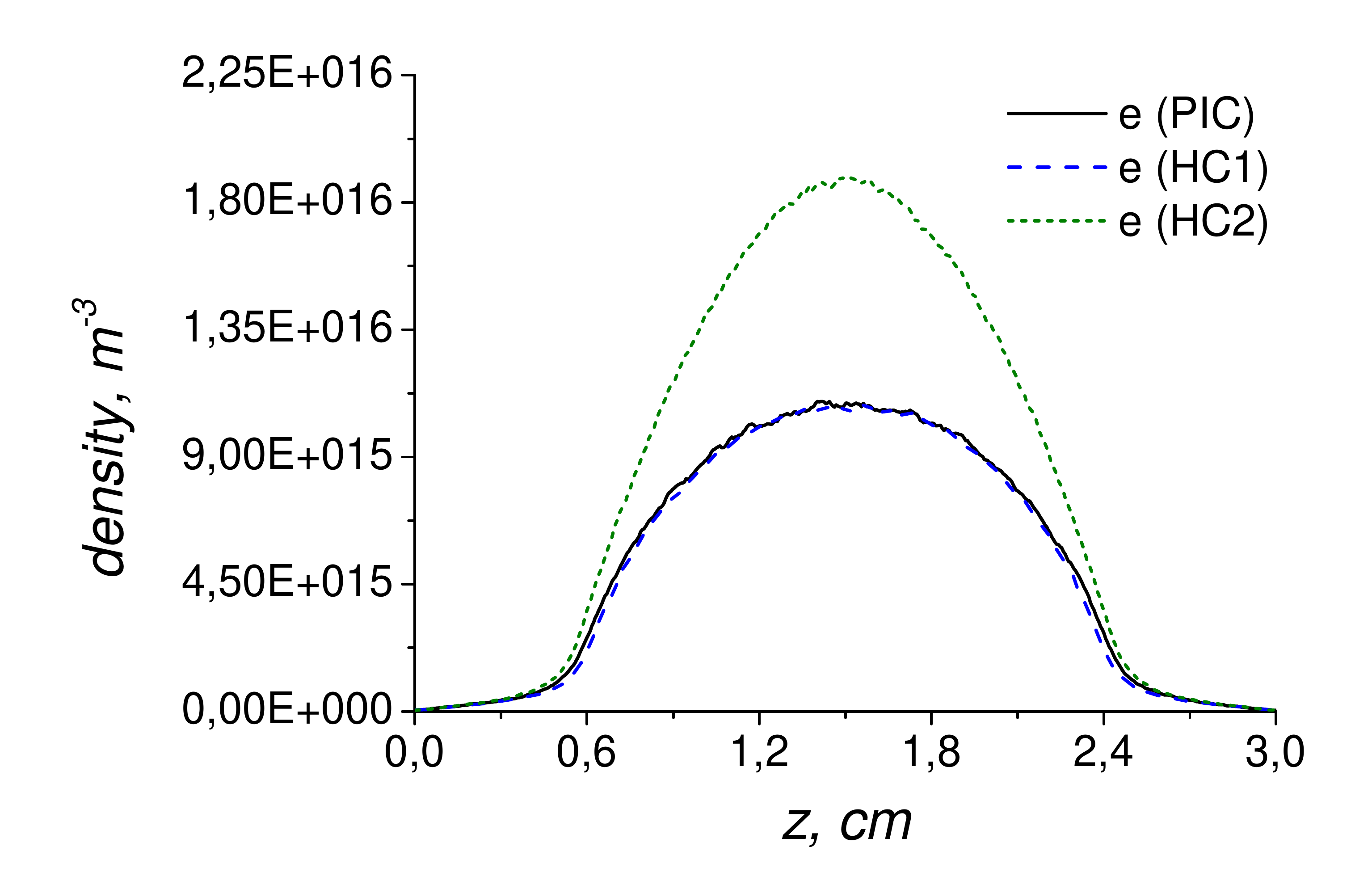}
\caption{The RF period-averaged density profiles for the Case IV ($\Omega (\alpha)$ mode, 300 mTorr) obtained by the PIC/MCC code and different versions of the hybrid code, see text.}
\end{figure}

Finally, to test how accurate are the ion fluid models at lower pressures we have studied an example of a pure He discharge operated at 300 mTorr pressure of
the He working gas,  $P_0/S = 170 \,\, {\rm W/m^2}$ and $l=3\,\,{\rm cm}$. The corresponding momentum exchange frequencies and mobilities to be used in the ion
fluid models were calculated with the Monte-Carlo code described before and fit with analytic functions as shown in Tables~\ref{t6} and \ref{t7}. At such a pressure
the momentum exchange frequency becomes comparable to the driving frequency and one can expect that the explicit ion flux time modulation on the  left hand side in Eq.~(\ref{eq2.2}) becomes comparable with the right hand side. This should lead to a breakdown of the drift-diffusion approximation. Indeed, the corresponding results (see Fig.~4) demonstrate that even the $\Omega$ mode (more commonly referred to as the $\alpha$ mode at such pressure) is not properly simulated by the hybrid code
using the drift-diffusion simulation with a realistic mobility (HC2). In contrast, the hybrid code version utilizing the full ion fluid model is in an excellent agreement with the kinetic result. Therefore, despite the drift-diffusion model is frequently used in the literature also at low pressures, one must be very careful in interpreting its results, 
as its assumptions are very likely to be violated there.

%One can see that whereas the hybrid code exploiting the full fluid model yields electron
%density being very close to the result of the PIC/MCC code, both drift-diffusion versions of the hybrid code provide significantly higher densities,
%suggesting that the ion inertia reduces plasma heating in this case as well. 

%%%%%%%%%%%%%%%%%%%%%%%%%%%%%%%%%%%%%%%%%%%%%%%%%%%%%%%%%%%%%%%%%%%%%%

\section{Conclusions}

%%%%%%%%%%%%%%%%%%%%%%%%%%%%%%%%%%%%%%%%%%%%%%%%%%%%%%%%%%%%%%%%%%%%%%

The present work describes a hybrid numerical scheme that can be used for simulations of highly collisional discharges with a complex chemistry. 
The scheme uses a kinetic description for electrons based on the PIC/MCC method and considers several possible fluid models for description of ion species.
For the ''full'' fluid model accounting for the explicit time modulation of the ion flux a simple numerical scheme is proposed. Its results are confronted with
results of the purely kinetic PIC/MCC code and the popular drift-diffusion approach for several exemplary discharges in regimes of practical interest. It is demonstrated that the drift-diffusion model with the constant mobility performs well for plasma discharges under atmospheric pressure unless the discharge is operated deeply in the $\gamma$ regime with heavy ion species present. It is also shown that the drift-diffusion approximation breaks down at lower pressures, whereas the hybrid code
version with the ''full'' fluid model remains very close to predictions of the kinetic code.

It is worth noting that the proposed ion fluid model taking into account ion inertia imposes only a minor complication of the numerical algorithm compared to the popular numerical schemes using the drift-diffusion approximation, which consists in solving an additional equation for the ion flux. However, considering the substantial improvement in the accuracy of the physics description of the former, we suggest that it should generally be preferred against the latter,
at least as far as RF-driven discharges are concerned.

%%%%%%%%%%%%%%%%%%%%%%%%%%%%%%%%%%%%%%%%%%%%%%%%%%%%%%%%%%%%%%%%%%%%%%

\section*{Acknowledgments}

The authors gratefully acknowledge support by DFG (German Research Foundation) within the framework
of the Research Unit FOR 1123. 

%the Deutsche
%Forschungsgemeinschaft via the Sonderforschungsbereich SFB-TR 87.

%%%%%%%%%%%%%%%%%%%%%%%%%%%%%%%%%%%%%%%%%%%%%%%%%%%%%%%%%%%%%%%%%%%%%%

\end{document}